# Quantum Algorithm for a Convergent Series of Approximations towards the Exact Solution of the Lowest Eigenstates of a Hamiltonian

Zhiyong Zhang

Stanford Research Computing Center, Stanford University, Stanford, CA94305

zyzhang@stanford.edu

**Abstract**. We present quantum algorithms, for Hamiltonians of linear combinations of local unitary operators, for Hamiltonian matrix-vector products and for preconditioning with the inverse of shifted reduced Hamiltonian operator that contributes to the diagonal matrix elements only. The algorithms implement a convergent series of approximations towards the exact solution of the full CI (configuration interaction) problem. The algorithm scales with $O(m^5)$, with $m$ the number of one-electron orbitals in the case of molecular electronic structure calculations. Full CI results can be obtained with a scaling of $O(nm^5)$, with $n$ the number of electrons and a prefactor on the order of 10 to 20. With low orders of Hamiltonian matrix-vector products, a whole repertoire of approximations widely used in modern electronic structure theory, including various orders of perturbation theory and/or truncated CI at different orders of excitations can be implemented for quantum computing for both routine and benchmark results at chemical accuracies. The lowest order matrix-vector product with preconditioning, basically the second-order perturbation theory, is expected to be a leading algorithm for demonstrating quantum supremacy for Ab Initio simulations, one of the most anticipated real world applications. The algorithm is also applicable for the hybrid variational quantum eigensolver.

**Introduction**. Feynman first envisioned the power of quantum computing for the simulation of physical systems by quantum computers (Feynman, 1982). The mid-1990s ushered in several breakthroughs in quantum algorithms for solving real world problems. Shor discovered quantum algorithms for factoring and computing discrete logarithms in polynomial times (Shor, 1997). Grover discovered quantum search algorithms (Grover 1996) (Grover 1997) that provably optimal. Simulations of quantum systems, particularly for applications to chemistry and materials sciences, have been primary inspirations of and have largely driven the development of quantum computation. Lloyd showed that, for a Hamiltonian $H$ of a sum of local non-commuting terms, a universal quantum simulator can simulate the Hamiltonian (Lloyd, 1996). Aspuru-Guzik (Aspuru-Guzik, 2005) first demonstrated the application of quantum computing algorithms to full configuration interaction (FCI) problems (L. Thøgersen, 2004) in quantum chemistry. Ground state eigenvalues of the Hamiltonian can be obtained with Hamiltonian simulation (Dominic W. Berry G. A., 2007) (Childs, 2010) by evolution of the system through the operator $e^{-H\tau}$, with (imaginary) time $\tau$, potentially approaching $\infty$, followed by quantum phase estimation, starting from an initial quantum state with non-exponentially small overlap with the ground state wavefunction. The operator $U = e^{-H\tau}$ needs to be approximated, and several methods, including product algorithm (Lloyd, 1996) (Dominic W. Berry G. A., 2007), Taylor series algorithm (Berry DW, 2015), Quantum Walk (Szegedy 2004) (Szegedy 2004) (Berry 2014) and Signal Processing and Qubitization algorithms (Chuang G. H., 2017) (Low & Chuang, 2019). Recently it was shown that an algorithm based on linear combinations of quantum walk steps of different orders, with expansion coefficients related to Bessel functions, can approximate the Hamiltonian evolution with linear $\Omega(t)$ and logarithmic $\Omega(\frac{\log 1/\varepsilon}{\log\log 1/\varepsilon})$ dependences that are (nearly) optimal with respect to the simulation time $t$ and the inverse of the desired error ε (D.W. Berry 2014). Clearly, the accuracy depends on the lengths of time, which in principle could be infinity, used for the Hamiltonian simulation as well as that for the phase estimation steps.

Single-particle approximations followed by electron correlations have long been the foundation of modern electronic structure theory. Flexible and highly correlated CI wavefunctions are arguably the most accurate and the most applicable, often the only reliable, electronic structure methods for a whole spectrum of applications, particularly for exceptionally reliable potential energy surfaces and excited states. The CI wavefunctions are expressed as linear combinations of configuration state functions (CSF) obtained from distribution of the **N** electrons in the one-particle basis state functions. Full CI includes all physically possible N-electron CSFs and solves the Schrodinger equation exactly for the given one-particle basis. The dimension of the full CI expansion grows exponentially with the size of the one-particle basis and the number of electrons though. The exponentially large Hamiltonian matrix and state vectors have mostly limited the CI methods to expansions truncated at singles and doubles excitations.

With quantum computing, the state vectors are represented compactly with qubits of a number that is polynomial in the size of single-particle basis, in contrast to classical representation of the wavefunction that is exponential in size. In this work we present a novel quantum algorithm for representing and optimizing the CI wavefunctions based on the direct application of the Hamiltonian on state vectors. The application of the Hamiltonian on the state vectors, or the Hamiltonian matrix-vector products, can be computed with polynomial scaling with the number of qubits and the number of terms in the Hamiltonian, as described later in detail. Together with an efficient quantum algorithm for preconditioning with the inverse of a shifted reduced Hamiltonian operator that contributes to the diagonal matrix elements only, a subspace of vectors of full CI dimensions can be formed and the subspace can be diagonalized to obtain the lowest eigenstates of the full CI of the Hamiltonian. The matrix-vector product with the initial guess vector of the zero order approximation, preconditioned with the inverse of the shifted matrix of diagonal elements, effectively gives rise to the second-order wavefunction in perturbation theory. Repeated applications of matrix-vector products and preconditioning effectively produce a convergent series of wavefunctions of higher order perturbations, and thus a whole range of flexible truncated wavefunctions that correspond to various orders of perturbation theory and/or CI expansions of various orders of excitations, all the way up to full CI, can be efficiently implemented in quantum computing. The preconditioned lowest order matrix-vector product then is expected be the most viable candidate for quantum supremacy for real applications, as the second order perturbation wavefunctions have been widely accepted as the simplest and yet often accurate enough method of all correlation methods in modern electronic structure theory.

**Davidson Iterative Subspace Methods**. Davidson subspace method (Davidson E. R., 1975) (Davidson E. , 1990), a cornerstone in modern CI methods, iteratively solves the large scale CI problem for the few lowest eigenstates and typically convergences in 10–20 iterations in practice. Two components of the Davidson method are the matrix-vector products and preconditioning with the inverse of the shifted matrix of the diagonal elements only. Consider the Hamiltonian of a generic interacting system in second quantized form,

$$H = \sum_{ij} v_{ij} a_i^+ a_j + \frac{1}{2} \sum_{ijkl} v_{ijkl} a_i^+ a_j^+ a_k a_l. \tag{1}$$

The $a_i^+(a_i)$ are, for example, the creation (annihilation) operators in Fermionic or bosonic systems. For electronic systems, the one-body and two-body interactions are the one-electron and two-electron integrals given by:

$$v_{ij} = h_{ij} = \langle i|t|j \rangle \text{ and } v_{ijkl} = [ij, kl] = \left\langle i(1)k(2) \left| \frac{1}{r_{12}} \right| j(1)l(2) \right\rangle. \tag{2}$$

The explicit spin indexes σ, τ are dropped for convenience for Hamiltonians without explicit consideration of spin-orbit interactions. However, the discussions in this work are applicable in situations with explicit consideration of spin-orbit interactions with addition of the spin-orbit interaction terms to the Hamiltonian.

The eigenvectors are expanded in a subspace of orthonormal vectors of reduced dimensions,

$$|x^k\rangle = \sum_{i=1,L} a_i^k |b^i\rangle, \tag{3}$$

where $a_i^k$ are the expansion coefficients and $|b^j\rangle$ is the *i*th expansion vector. The Hamiltonian is represented by the expansion vectors and new expansion vectors are formed from products of the Hamiltonian and the expansion vectors, $\sigma_j = H|b^j\rangle$. During each iteration, the Hamiltonian is diagonalized within the subspace to obtain approximations to the true eigenvalues and eigenvectors. When $|x^k\rangle = |c^k\rangle$ is exact, the Hamiltonian matrix-eigenvector product, $H|x^k\rangle$, satisfies $(H - \lambda^k)|x^k\rangle = 0$, where $\lambda^k$ is the *k*th eigenvalue of the exact solution. The corrections $|\delta^k\rangle = |c^k\rangle - |x^k\rangle$ to the approximate eigenvectors $|x^k\rangle$ can be related to the residual vector $|r^k\rangle = (H - \lambda^k)|x^k\rangle$:

$$(H - \lambda^k)|\delta^k\rangle = -(H - \lambda^k)|x^k\rangle = -|r^k\rangle \tag{4}$$

The Davidson algorithm adds $|\delta^k\rangle$ as new expansions vectors to the subspace. (Davidson E. R. 1975) (Liu, 1978) (MATTHEW L. LEININGER, 2001) The Davidson approach forms the bedrock for the practical solution of large Hamiltonian matrices in quantum chemistry. The success of the Davidson approach largely derives from the judicious choices of the update vectors obtained from the above equation. It can be shown with spectral decomposition that $|\delta^k\rangle$ computed from the above equation selectively amplifies those components of the residual vector that are in proximity and contribute most significantly to the eigenvectors sought after. In practice, the exact eigenvalues $\lambda^k$ are replaced with the approximate eigenvalues $\rho^k$ from the diagonalization of the

Davidson subspaces, which is shown to be equivalent to the Newton-Raphson updates for finding the stationary points of the Rayleigh quotients,

$$\frac{\langle x^k|H|x^k\rangle}{\langle x^k|x^k\rangle} \tag{5}$$

The bottlenecks of the Davidson procedure are the computation of the Hamiltonian matrix-vector products for the computation of the update vectors of the subspace expansion and for the calculation of matrix elements of the Hamiltonian in the Davidson subspace. In the traditional CI methods, the dimensions $N$ of the solution eigenvectors in the Hilbert space scales exponentially with the size of the system in terms of number of one-electron orbitals and number of electrons. In quantum computing, the resulting vectors of the Hamiltonian matrix-vector products could be encoded as a state vector of the quantum system that encodes the computation and the procedure will be presented below, thus eliminating the biggest obstacle in the traditional CI problems.

Critically, construction of the update vectors $|\boldsymbol{\delta}^k\rangle$ in a way that captures the optimal convergence towards the exact solution in a series of expanded subspace representations is of paramount importance to the success of the Davidson approach. The residual vectors, obtained from the matrix vector products, need to be conditioned by solving equation x. Enormous amount of research has been devoted to precondition techniques that accelerate convergence properties of iterative approaches in traditional and quantum computing. Exact solution of equation x involves the inversion of the shifted Hamiltonian $H - \rho^k$, again with the unknown exact eigenvalue replaced by the approximate eigenvalues. The exact inversion involves the solvation of the Hamiltonian eigenvalue problem itself. Instead, approximate precondition methods have been shown efficient alternatives that resulted in rapid convergence of the approximate solutions to the exact values within given precisions. It has been very well established from experiences in traditional CI problems that it only requires a number of iterations on the order of 10 to 20 for the approximation to reach an accuracy of $10^{-8}$. The most widely used approximation is to replace the Hamiltonian $H$ in equation x with its diagonal part $D$,

$$|\boldsymbol{\delta}^k\rangle \cong -(H - \rho^k)^{-1}(H - \rho^k)|x^k\rangle \cong -(H_D - \rho^k)^{-1}|r^k\rangle \tag{6}$$

**Quantum Algorithm for Hamiltonian Matrix Vector Product**. For a Hermitian Hamiltonian of a sum of unitary operators $U_i$, $H = \sum_{i=1}^{M} \alpha_i U_i$, its application on a quantum state, $H|\Psi\rangle$, can be obtained probabilistically. For simplicity, we assume that the number of terms in the expansion satisfies $M = 2^m$. Without loss of generality, it is assumed that $\alpha_i \geq 0$, as extra signs can be absorbed into the definition of the unitaries $U_i$ and that the positive coefficients can be normalized with $a = \sum_{i=1}^{M} \alpha_i$. The register of $n$ qubits encoding the quantum state $|\Psi\rangle$ is augmented with a register of $m$ ancilla qubits to index and control the individual unitaries in the expansion:

$$W = \sum_{i=0}^{M-1} |i\rangle\langle i| \otimes U_i \qquad (7)$$

With a unitary operator on the ancilla register that satisfies

$$V|0^m\rangle = \frac{1}{a}\sum_{i=0}^{(M=2^m)-1} \sqrt{\alpha_i}|i\rangle \qquad (8)$$

we can construct a pure quantum state in the Hilbert space of the $m + n$ qubits, with the desired quantum state $H|\Psi\rangle$ given by the components indexed by the $|0^m\rangle$ component of the ancilla register:

$$V^+WV(|0^m\rangle \otimes |\Psi\rangle) = |0^m\rangle \otimes |\Phi\rangle + (|0^m\rangle \otimes |\Phi\rangle)_\perp \qquad (9)$$

with $|\Phi\rangle = H|\Psi\rangle = \sum_{i=0}^{M-1} \alpha_i U_i |\Psi\rangle$. The second term of the above equation with the subscript $\perp$ is the state function that is orthogonal to the first term. Thus, the projection of the first register of $m$ ancilla qubits to the $|0^m\rangle$ component will leave the state $H|\Psi\rangle$ in the second register of the $n$ qubits for the state vectors.

**Quantum Algorithm for Preconditioning with Shifted Inverse Diagonal Hamiltonian Matrix.** Paradoxically, the advantage of replacing the exact Hamiltonian with its diagonal counterpart is not obvious. The diagonal Hamiltonian matrix is still of dimension $N$. That is not a problem in the case of the traditional approach, as vectors of the whole dimensionality $N$ are explicitly represented with each component calculated and stored. Apparently, straightforward application of equation $|\delta^k\rangle \cong -(H - \rho^k)^{-1}(H - \rho^k)|x^k\rangle \cong -(H_D - \rho^k)^{-1}|r^k\rangle$

(6 is not going to work and a completely different approach for the calculation and representation of the preconditioned residual vectors in the case of quantum computing is needed for the successful application of the Davidson diagonalization procedure.

For the following discussion, we consider the electronic Hamiltonian within the Graphical Unitary Group Approach (GUGA). (Werner Dobrautz, 2019) The Hamiltonian is a sum of one-body and two-body terms:

$$H = H_1 + H_2 = \sum_{ij} t_{ij} E_{ij} + \frac{1}{2} \sum_{ijkl} v_{ijkl} e_{ij,kl} \qquad (10)$$

The one-body and two-body excitation operators are defined as

$$E_{ij} = \sum_\sigma a_{i\sigma}^+ a_{j\sigma} \qquad (11)$$

$$e_{ij,kl} = \sum_{\sigma\tau} a_{i\sigma}^+ a_{k\tau}^+ a_{l\tau} a_{j\sigma} = E_{ij}E_{kl} - \delta_{jk}E_{il} \qquad (12)$$

and the summation is over the spin indexes. These operators preserve the spin of the system. We consider the part of the Hamiltonian $H_D$ that contributes to the diagonal matrix elements, given as

$$H_D = \sum_i t_{ii} E_{ii} + \frac{1}{2} \sum_i v_{iiii}(E_{ii}^2 - E_{ii}) + \frac{1}{2} \sum_{i\neq j} v_{iijj}(E_{ii}E_{jj}) + \frac{1}{2} \sum_{i\neq j} v_{ijji}(E_{ij}E_{ji} - E_{ii}) \qquad (13)$$

In the above, all the terms, except for the terms of $E_{ij}E_{ji}$, contribute to the diagonal matrix and the diagonal matrix elements only. When the two orbitals $i$ and $j$ in the raising and lowering operators in $E_{ij}E_{ji}$ are both singly occupied but with opposite spins, care must be taken that only results that lead to the original occupancy configurations after the application of the operator $E_{ij}E_{ji}$ are kept when computing the diagonal matrix elements.

With the Hamiltonian operator $H_D$ for the calculation of the diagonal matrix elements, the inversion of shifted diagonal Hamiltonian matrix could be efficiently implemented to precondition the residual vector with a quantum algorithm that implements the linear combinations of unitaries for the shifted diagonal Hamiltonian $(H_D - \rho^k)^{-1}|r^k\rangle$. In the flowing discussion we omit the explicit inclusion of the shift $\rho^k$ and instead the following discussion only refers to $H_D^{-1}|r^k\rangle$, as $\rho^k$ is just a constant. Like the algorithm for the matrix vector product considered earlier, we use a register of $m$ ancilla qubits that is large enough for the controlled operation of the individual terms in the diagonal Hamiltonian matrix. Then with $W = \sum_{i=0}^{M-1}|i\rangle\langle i| \otimes U_i$ and $V|0^m\rangle = \frac{1}{a}\sum_{i=0}^{(M=2^m)-1}\sqrt{\alpha_i}|i\rangle$, as constructed accordingly for the diagonal Hamiltonian, we have

$$V^+WV(|0^m\rangle \otimes |r^k\rangle) = |0^m\rangle \otimes |H_D' r^k\rangle + \left(|0^m\rangle \otimes \left|H_D' r^k\right\rangle\right)_\perp \quad (14)$$

$$= |0^m\rangle \otimes \sum_{j=0}^{N-1} c_j^k H_D'|\varphi_j\rangle + \left(|0^m\rangle \otimes \left|H_D' r^k\right\rangle\right)_\perp \quad (15)$$

$$= |0^m\rangle \otimes \sum_{j=0}^{N-1} c_j^k \xi_j|\varphi_j\rangle + \left(|0^m\rangle \otimes \left|H_D' r^k\right\rangle\right)_\perp \quad (16)$$

Here, $c_j^k$ and $\xi_j$ are the expansion coefficients of the residual vector $|r^k\rangle$ in the basis of $|\varphi_j\rangle$ and the corresponding diagonal Hamiltonian matrix elements in the basis of $|\varphi_j\rangle$. In the same spirit as in (A. W. Harrow, 2009), and conditioned upon the control register being in the state $|0^m\rangle$, an extra ancilla qubit $|0\rangle_a$ could be added and rotated to the state $\frac{C}{\xi_j^2}|0\rangle_a + \frac{C}{\sqrt{1-\xi_j^2}}|1\rangle_a$, where $C$ is a normalization factor, we obtain

$$|0^m\rangle \otimes \sum_{j=0}^{N-1} c_j^k \xi_j|\varphi_j\rangle \otimes (\frac{C}{\xi_j^2}|0\rangle_a + \frac{C}{\sqrt{1-\xi_j^2}}|1\rangle_a) + \left(|0^m\rangle \otimes \left|H_D' r^k\right\rangle\right)_\perp \otimes |0\rangle_a \quad (17)$$

Finally, with projections to the $|0^m\rangle \otimes |0\rangle_a$ state of the control register and the rotation ancilla qubit, we are left with the desired residual vector pre-conditioned with the inversion of the diagonal matrix Hamiltonian matrix shifted with the approximate eigenvalues of the $kth$ iteration of the solution of the subspace, $|r^k\rangle_{precond} = (H_D - \rho^k)^{-1}|r^k\rangle$.

Preconditioning plays a critical role in the accelerated convergence of the subspace iterative diagonalization approach. The Hamiltonian matrix vector products are performed repeatedly at each iteration to test if the solution is converged to the desired precision and are used to compute the residual vectors that are preconditioned before added to the subspace as additional expansion vectors. Preconditioning with the inverse of the shifted diagonal Hamiltonian matrix has been successfully applied in the Davidson approach for the CI problem and has been critical for the wide application of the classical configuration interaction approaches, partly due to the simplicity in the inversion of the diagonal matrix within the classical approach. There have been large bodies of work on different precondition techniques for CI and other applications involving large matrices. The precondition amounts to rotations of the subspace spanned by vectors computed from the matrix vector products. Intuitively, the precondition rotates the subspace to better align with the subspace spanned by the actual eigenvectors of the Hamiltonian, thus accelerate the convergence of the solution. One of the most simple and effective preconditioning is formally captured by the inverse of the shifted Hamiltonian as formally expressed in $|\delta^k\rangle \cong -(H-\rho^k)^{-1}(H-\rho^k)|x^k\rangle \cong -(H_D-\rho^k)^{-1}|r^k\rangle$ (6).

In general, efficient inversion algorithms, as a special case of action of a smooth function of the Hermitian matrix on a quantum states, can be used to implement the preconditioning. A recent work (Sathyawageeswar Subramanian, 2019) has classified the implementation of the application of Hermitian matrices on quantum states into three broad categories: (1) Hamiltonian simulation and Quantum Phase Estimation (QPE) for representing the results in the spectral basis of the matrix (A. W. Harrow, 2009) (Prakash); (2) using Linear Combinations of Unitaries (LCU) (Wiebe, 2012) (D. W. Berry, 2015) (A. M. Childs, 2017) to embed the resultant states in a larger state space with auxiliary qubits; and (3) Qubitization and Quantum Signal Processing (QSP) that uses a signal state for the action of the Hermitian matrices on the input states. (Chuang G. H., 2017) (A. Gilyén). As described in (Kyriienko, 2020) for the quantum inverse iteration algorithm, inversion of the Hermitian matrix can be implemented as a linear combination of Hermitian matrices to different powers and implemented with a variety of methods described in (Sathyawageeswar Subramanian, 2019). In addition, inversion free technologies for preconditioning (YE)could also be explored for quantum computing implementations.

**Scaling of The Algorithm.** In the case of electronic structure calculations, with a mean-field one-electron approximation, such as in the self-consistent field theory, as the starting approximation, the matrix vector product captures the effects of singles and doubles excitations. The order of matrix vector products required to reach full CI expansion for the wavefunction would be $K = O\left(\frac{n}{2}\right)$, or just $\sim O(n)$, where the number of electrons is $n$. The complexity of implementing the $Kth$ order matrix vector products $H^K|x\rangle$, neglecting the interleaving preconditioning steps, would be $K$ times the cost of implementing a matrix product of the Hamiltonian. The complexity of implementing a single matrix vector product scales with the logarithm of the number of terms in the Hamiltonian, as determined by the number of auxiliary qubits needed for the controlled application of individual terms of the Hamiltonian on the state qubits. The controlled operations can be implemented with a control that indexes the Hamiltonian matrix product and $O(M(m + \log M))$ (Dominic W. Berry A. M., 2015) (D.W. Berry 2014) gates for the controlled implementation of the individual terms in the Hamiltonian. Notice that the total number of terms

$M$ in the Hamiltonian is on the order of $m^4$ for an electronic structure Hamiltonian, and we arrive at an overall scaling of $O(nm^5)$, where $n$ and $m$ are the number of electrons and the number of one-electron orbitals, respectively.

The multiplicative factor for the scaling is largely determined by the number of iterative iterations needed to converge the solution to the desired accuracy. As in the Davison approaches in traditional quantum chemistry CI algorithms, preconditioning could be the key to the accelerated convergence of the iterative solutions. When it is possible to project the Hamiltonian to a reduced Hamiltonian that contributes to the diagonal elements only, as assumed in this work, it is possible to implement an efficient quantum algorithm for the inverse of the shifted diagonal matrix for preconditioning the update vectors to be added to the subspace. In general, the specific approaches and quantum algorithms for preconditioning are still wide open for research, just as in the classical approaches for preconditioning (Andreas Stathopoulos, 1995) (Yunfeng Cai, 2013) (Davidson E. , 1990). Several recent progresses in quantum algorithms for preconditioning can be found in (Benzi 2002) (B. D. Clader, 2013) (Changpeng Shao, 2018) (Jonathan Welch, 2014). We note that in general the diagonal matrix itself is of the dimension of the Hamiltonian matrix and preconditioning based on that could scale as the order of the Hamiltonian matrix, if there exist not a diagonal Hamiltonian that contribute to the diagonal elements only. In that case it is still possible to precondition the wave vectors with a diagonal matrix of reduced dimensions, for example, from a space of singles and doubles excitations of some reference vectors. As often noted in the case of Davison approach in the classical case, the number of iterations required to converge the sunspace solution to an accuracy in energy of $10^{-8}$ in 10 to 20 iterations. It is reasonable to expect similar number of iterations may be enough to converge the solution of the subspace problem with the quantum algorithm and we expect an overall scaling of the algorithm in the order of of $O(nm^5)$, with a prefactor of 20, depending on the convergence pattern of the specific problems.

**Combination of Subspace Algorithm with VQE**. The hybrid classical quantum variational quantum eigensolver (VQE) algorithm (Alberto Peruzzo, 2014) (Jarrod R McClean, 2016) has recently emerged as a promising algorithm for quantum simulation of chemical systems. VQE uses a parameterized wavefunction ansatz with a set of few parameters. In the hybrid algorithm, the ansatz is encoded on a quantum system to obtain the expectation values of the Hamiltonian (eigenvalues) as a functional of the variational parameters and classical algorithms are used to determine the updated ansatz for evaluation on the quantum system. Since the algorithm only requires the evaluation of expectation energies of the Hamiltonian, the depth of the quantum circuit is minimal and thus well suited for the near-term quantum computers with noisy qubits and short entanglement time. However, a severe limitation of VQE is that the accuracy for the wavefunction and eigenvalue obtainable is dictated by the form of the parameterized variational function. The subspace algorithm developed in this work can be applied to any initial guess vectors that can be encoded on the quantum device. In fact, the choices of the initial guess vectors are important considerations for the rapid convergence to the desired solutions in our algorithm. Application of the subspace algorithm to the VQS wavefunction with converged parameters provide a straightforward and powerful mechanism for eliminating the limitations imposed by a parameterized ansatz of fixed form. Since the subspace algorithm captures the corrections due to the part of the Hilbert space that is orthogonal to the approximate wavefunction with maximal overlap with the true wavefunction facilitated by the preconditioner, we expect that judicious choice of the VQE wavefunctions as the initial guess vectors for the subspace algorithm could be the methods of choice for simulations on the near-term and future quantum devices.

The two sets of parameters of the SQE-VQE algorithm provide rich possibilities for efficient implementation. For the near-term devices, one possibility is to limit the subspace algorithm to one single iteration. The subspace will consist of the VQE wavefunction and the update vector computed from the product of the Hamiltonian with the vector of the VQE wavefunction. One possibility is to use the converged VQE wavefunction and then solve the two-dimensional subspace problem. It is also conceivable that we can use a wavefunction that is extended with the subspace solution at each iteration for the variational determination of the VQE parameters. One single step of matrix vector product is expected to significantly improve the accuracy. As the coherence time of the entanglement improves, it is easily foreseeable that further steps of Hamiltonian vector products can be incorporated for systematic improvements of accuracy.

**Discussions and Conclusions**.

Formally the exact solution of the Schrodinger equation could be obtained with the time-evolution techniques discussed in the introduction. For practical applications, truncation of the time evolution with discretized time steps has been the standard approach for the solution of Hamiltonian problems. Ultimately the success of the algorithm for practical applications depends on the efficient convergence to the solution of desired accuracy. However, the longtime steps necessary for the convergence has limited its practical applications. Currently, the most popular VQE approach, though widely applied in recent work, is severely limited with the forms of the wavefunction that could be encoded with hardware parameters.

The algorithm we presented with quantum algorithms for matrix-vector products and preconditioning essentially defines a convergent series of approximations all the way up to the exact solution of the exact full CI problem for the Schrodinger equation at a given level of truncation for the one-particle basis functions. The lowest order of approximation with one order of matrix-vector product, preconditioned with the inverse of the shifted Hamiltonian matrix of diagonal elements, is essentially the second order perturbation approximation that have long been the simplest but yet physically meaningful approximations in electronic structure theories. To demonstrate quantum supremacy (Arute, 2019) (Boixo, 2018) (al., 2018) (Andrew M. Childs, 2018)for real word applications, particularly of Ab Initio simulations in chemistry and material sciences, the approach we provided here is no doubt the leading candidate. Low orders of matrix-vector products, properly preconditioned and with/without the diagonalization of the subspaces of approximation vectors, essentially implement the equivalents of the widely used approximations for electron correlation method in quantum chemistry, including but not limited to perturbation and/or CI of up to quadruple excitations that are widely believed to give rise to accurate benchmark results at chemical accuracy (III, 1999) (Raghavachari, 1987) (Krishnan Raghavachari, 1989). We foresee that further development and implementation of the algorithm presented in this work could lead to a standard approach for simulating quantum systems, especially the molecular systems, which could become the foundation for the practical applications of quantum computing to real world problems for a wide range of applications in chemistry, materials sciences, and biological sciences and so on.

Our algorithm, formally scaling as $O(nm^5)$ as discussed previously, likely is optimal in the convergence towards the exact solutions, as justified by the variational convergence property of the Ritz subspace. Subspace methods, particularly as formulated by Davidson, formed the foundation of the practical applications of large-scale configuration interaction (CI), arguably the most flexible and most accurate approaches in quantum chemistry. Due to the prohibitive exponential scaling of full CI with the size of the problem, only truncated CI spaces with singles and doubles excitations from reference vectors are feasible for larger problems. With a truncated CI expansion, size consistency and size extensivity have been notorious problems that often need to be reckoned with (Bartlett, 1981). The quantum algorithms developed in this work, when formally approaches exact solutions, could be truncated at desired accuracies and/or, alternatively, at specific orders of Hamiltonian matrix-vector products. The iterative solutions involve summations of different orders of matrix vector products. The update vectors can be limited to summation to a maximum order of matrix vector product. For example, if we limit the update vector to only the single order matrix vector product, essentially, we are limiting the solution to singles and doubles excitations relative to the reference vector or vectors. With limitations to second order matrix vector products, we expect the quality of the solution to be comparable to that of CI with up to quadruple excitations, approximations that often rival those of the near gold standard results of the widely used coupled cluster methods (Cremer, 2013).

The algorithms developed in this work is general and flexible in several respects, either in terms of the order of matrix vector products that are related to the excitation levels, or in terms of the initial guess vectors, of which the recently developed hybrid variational quantum eigen solvers could be easily incorporated and extended with desired order of matrix vector products for much improved accuracy and flexibility. As pointed out earlier, even an extension with one iteration of iterative space updates to the VQE approach, coupled with an efficient variational guess vector that could be encoded on the hardware, could greatly enhance the accuracy of the results. The approach could be ideally suited for the intermediate quantum computing platforms with limited coherence time for the qubits and at the same time easily applicable to quantum computing platforms with increased number of qubits and coherence times. Immediate further work may include the implementation of the algorithm developed here on a real quantum computing platform. Equally important, studies of the convergence properties in terms of the number of iterations, order of matrix vector products, and particularly the effects of the preconditioning on the convergence properties on full CI problems on classical computing platforms could shed more light and help improve the implementation and further development of the proposed algorithm.

**Acknowledgement**. The author gratefully acknowledges the insightful comments and discussions by Dr. Travis L. Scholten during the preparation of this manuscript.